\documentstyle[12pt,subeqnarray]{article}
\def\be{\begin{equation}}
\def\ee{\end{equation}}
\def\beq{\begin{eqnarray}}
\def\eeq{\end{eqnarray}}
\def\lsim{\:\raisebox{-0.5ex}{$\stackrel{\textstyle<}{\sim}$}\:}

\begin{document}
\begin{flushright}
Fermilab Pub-97/219-T\\
TIFR/TH/97-29 \\
hep-ph/yymmnn\\
July 1997
\end{flushright}
\bigskip
\begin{center}
{\Large{\bf Implications of the HERA Events for the R-Parity Breaking
SUSY Signals at Tevatron}} \\[3cm]
{\large Monoranjan Guchait$^1$ and D.P. Roy$^{1,2}$} \\[1cm]
$^1$ Tata Institute of Fundamental Research, Homi Bhabha Road, Bombay
(Mumbai) 400 005, India. \\ [2mm]
$^2$ Fermi National Accelerator Laboratory, P.O. Box 500, Batavia,
Illinois 60510, USA.\\
\end{center}
\bigskip\bigskip
\begin{center}
\underbar{\bf Abstract}
\end{center}
\medskip

The favoured R-parity violating SUSY scenarios for the anomalous HERA
events correspond to top and charm squark production via the 
$\lambda'_{131}$ and $\lambda'_{121}$
couplings. In both cases the corresponding electronic branching
fractions of the squarks are expected to be $\ll 1$. Consequently the
canonical leptoquark signature is incapable of probing these scenarios
at the Tevatron collider over most of the MSSM parameter space. We
suggest alternative signatures for probing them at Tevatron, which
seem to be viable over the entire range of MSSM parameters.

\newpage

\noindent{\bf 1.~~ \underbar{The Anomalous HERA Events}} :\\

The H1 and ZEUS experiments have reported some anomalous high-Q$^2$
events from the HERA $e ^+p$ collider, which could be suggestive of
new physics beyond the Standard Model (SM). The H1 experiment has
reportedly seen 12 neutral-current events, $e ^+p \rightarrow e ^+qX$,
at $Q^2 >$ 15,000 GeV$^2$ against the SM prediction of 5 [1]; while
ZEUS has reported 5 events at $Q^2 >$ 20,000 GeV$^2$ against the
prediction of 2 [2]. Moreover the excess of 7 events observed by H1
seem to cluster around a common $e ^+q$ mass of
\be
M \simeq 200 GeV ,
\ee
which is not inconsistent with those of the ZEUS events [3]. Taken
together, they represent an excess of 10 events, with a common mass
range of 200--220 GeV. They are based on the 1994--'96 data,
corresponding to a combined luminosity of 34 $pb^{-1}$ for the two
experiments, while the reported detection efficiency for each
experiment is about 80\%. This corresponds to a cross-section of
\be
\sigma \simeq 0.4 pb
\ee
for these anomalous events.

These events have aroused a good deal of excitement in high energy
physics in the past few months; and several possible mechanisms of new
physics have been suggested [4-10]. It should be noted of course that
the statistical significance of the signal is about $3\sigma$ for each
experiment, which is by no means conclusive. The ongoing experiments at
HERA are expected to double the data sample in another year. While
very welcome, this may not be sufficient to settle the issue
conclusively. It is imperative therefore to see if the contending new
physics mechanisms can be tested at other colliders --- in particular
at Tevatron, which has a considerable energy reach to probe these 
mechanisms.

The new physics mechanisms suggested are mainly of two types --- (i) a
contact interaction term corresponding e.g. to a heavy $Z^{\prime}$ exchange
[6,8,9], and (ii) the production and decay of a generic leptoquark,
i.e. a hypothetical particle coupling to the $e ^+q$ channel [4-8].
The first interpretation seems to be disfavoured on several grounds.
The size of the contact interaction term required is at best
marginally compatible with the upper limits from LEP and Tevatron
colliders. Moreover it favours the standard $e ^+q$ mass distribution,
$M = \sqrt{s \cdot x}$, as given by the $x$ distribution of the quark,
instead of its clustering at a high value of $M$. For the same
reason, the $M$ integrated cross-section is expected to go down at
large $Q^2$. Instead, the clustering in $M$ and the flat distribution
over a very large range of $Q^2$ observed in the data clearly favour
the formation and decay of a generic leptoquark in the $e ^+q$
channel. 

Thus, it is natural to ask whether some of the extensions of the SM
can naturally account for such a leptoquark. The leptons and quarks are
unified in GUT, which naturally predict leptoquark states both as
gauge (vector) bosons and Higgs scalars [10, 11]. However, the
exchange of these objects in GUT generally leads to lepton and baryon
number violating interactions, and in particular to proton decay. Thus
the stability of proton implies these objects to be very heavy $(M >
10^{15}$ GeV), which puts them far beyond the reach of present or
foreseeable future machines. While there are examples of GUT models
like $E_6$, having leptoquarks without baryon number violating
couplings [10], there is no natural reason to expect them to be as
light as a few hundred GeV. 

A more plausible candidate for a generic leptoquark in the mass range
of a few hundred GeV is the scalar superpartner of quark (squark) in
the R-parity violating SUSY model [12, 13]. In this case they can have
lepton and baryon number violating Yukawa couplings and mediate proton
decay as well. Unlike the gauge couplings, however, these Yukawa couplings are
not constrained by any symmetry consideration. Therefore one can
assume a finite value for the lepton number violating coupling while
setting the baryon number violating ones to zero. The former ensures
squark coupling to the $e ^+q$ channel while the latter prevents
proton decay. Thus in the R-parity violating SUSY model the squark can
masquerade as a leptoquark and naturally account for the anomalous
HERA events. Indeed these squarks are by far the most promising new
physics candidates for the HERA events; and as such they have
attracted a good deal of attention in the current literature on this
subject [4-7]. The purpose of this work is to identify the most
plausible R-parity violating SUSY scenarios for the anomalous HERA
events and study the corresponding signals for the Tevatron collider.

\vskip 2em

\noindent{\bf 2.~~ \underbar{R-Parity Breaking SUSY Scenarios}} :\\

We shall consider the minimal supersymmetric extension of the standard
model (MSSM) with explicit R-parity breaking [13]. The latter 
arises from the
following Yukawa interaction terms in the Lagrangian:
\be
L = \lambda_{ijk} l_i \tilde l_j \bar e_k + \lambda'_{ijk} l_i \tilde
q_j \bar d_k + \lambda''_{ijk} \bar d_i \tilde{\bar d}_j \bar u_k , 
\ee
plus analogous terms from the permutation of the tilde, denoting the
scalar superpartner. Here $l$ and $\bar e$ (q and $\bar u$, $\bar d$) 
are
the left-handed lepton doublet and antilepton singlet (quark doublet
and antiquark singlet) fields; and 
i, j, k are the generation indices. The terms relevant for the HERA
events are 
\be
\lambda'_{ijk} (l_i \tilde q_j \bar d_k + l_i q_j \tilde{\bar d}_k) + h.c. 
\ee

It is customary to assume a hierarchical structure for these Yukawa  
couplings in analogy with the standard Yukawa couplings of the quarks 
and leptons. The squark formation and decay processes corresponding 
to different choices of the leading $\lambda'$ coupling of eq.(4)
are shown in Table - I. Only the 1st row corresponds to squark 
formation from a valence quark, while all other cases are from 
sea quarks. Knowing these quark fluxes one can easily calculate 
the cross-section for these processes at HERA for a given $\lambda'
\sqrt B$, where B denotes the squark branching fraction into the 
shown channel. The 2nd column shows the sizes of corresponding
$\lambda' \sqrt B$, required to explain the cross-section (2) of the
HERA events [5,6]. While the required size of the quantity is
small for the valence quark case (1st row), it is larger by an
order of magnitude in the other cases. Note that for the last
two rows there is an equal probability of $\tilde{\bar d}_j$ decay
into the neutrino channel, i.e. $B \leq$ 1/2.

The last column shows the upper limits on these R  violating 
couplings from other processes. The $\lambda'_{111}$ limit
comes from neutrinoless double beta decay [14], $\lambda'_{112,113}$
limits from charge current universality [15], $\lambda'_{121,131}$
limits from atomic parity violation [16], $\lambda'_{122,133}$ limits
from $\nu_{e}$ mass [17], $\lambda'_{123}$ limit from forward-backward
asymmetry [15], and $\lambda'_{132}$ from Z-decay [18]. All but the
last of these limits are taken from the recent compilation of 
ref.[5].They are all 1$\sigma$ limits. 

More recently a precise measurement of atomic parity violation
in ${}^{55}Ce_{133}$ has been reported[19]. The measured value of the
weak charge is ${Q^{ex}}_W = -72.11 \pm 0.27 \pm 0.89$, where the 2nd
error is theoretical. This is in remarkable agreement with the SM
value of ${Q^{SM}}_W \simeq -72.9$[20]. From
\be
Q_W = -2[C_{1u} (2 Z+N) + C_{1d} (Z +2N)],
\ee
\beq
\triangle C_{1d} = \frac{{\lambda^{\prime}_{1j1}}^2 \sqrt 2}{8 
{M^2_{\tilde 
q_j}} G_F}
\eeq
it is clear that R-violating SUSY contribution can only add to the
magnitude of ${Q^{SM}}_W$. Consequently a 1$\sigma$ bound would 
imply a very severe restriction on the $\lambda'_{1j1}$ coupling.
But such a bound would not be justified since the data point is roughly
1$\sigma$ below the magnitude of SM. Therefore we have estimated the
2$\sigma$ bound, where the experimental and the theoretical errors
of ${Q^{ex}}_W$ have been added linearly. The resulting bound for
a 200 GeV squark, ${\lambda'}_{1j1} < 0.10$, is shown in the bracket
in Table - I. This bound is consistent with the one obtain in
ref.[21]. 

Comparing the 2nd and 3rd columns of this Table, we see 
that the required coupling is reasonably small compared to its 
upper limit for the charm and top squark productions from a
valence quark, i.e. 
\beq
&&e ^+d \rightarrow \tilde c_L \rightarrow e ^+d , \\ [2mm]
&&e ^+d \rightarrow \tilde t_L \rightarrow e ^+d ,
\eeq
where the subscript $L$ denotes left chirality. 
Moreover, the inclusion of NLO correction in eq.(7) and (8) is expected
to reduce the required value of $\lambda'_{1j1} \sqrt B$ further
by $\sim$ 30\%[22].
Therefore we shall
concentrate on these two cases, in studying the R violating SUSY
signal at the Tevatron. There is only one other case, where the 
required $\lambda'\sqrt B$ is compatible with the upper limit
on $\lambda'$. This corresponds to top squark production from the
strange quark sea,
\be
e ^+s \rightarrow \tilde t_L \rightarrow e ^+s .
\ee
This case has been recently studied in [18], where it was shown to 
give a consistent solution to the HERA anomaly over a limited 
range of the MSSM parameter space. We shall discuss this possibility
while studying the top squark production scenarios of eq.(8) at the 
Tevatron collider.

A brief discussion of the squark branching fraction B is in order.
Under the assumption of hierarchical $\lambda'$ couplings, there is 
only one dominant R violating channel in a given scenario. 
The corresponding 
decay width of squark is 
\be
\Gamma_{R\!\!\!\!/} = \frac{1}{16\pi} \lambda^{\prime 2} M.
\ee
In addition the squark has R conserving decays into chargino and 
neutralino channels. The corresponding decay width $\Gamma_R$
will be a function of the MSSM parameters, but independent  
of $\lambda'$. Thus 
\be
B = \frac{\Gamma_{R\!\!\!\!/}}{\Gamma_{R\!\!\!\!/} + \Gamma_R} .
\ee
Note that the product $\lambda'\sqrt B$ is constrained by the 
cross-section of the HERA events, as shown in the second column
of Table--I. This can be used to eliminate $\lambda'$ from eqs.(10,11),
so that in any given scenario the branching fraction B is a unique
function of the MSSM parameters. Thus  
\be
B = \frac{\sqrt{1 + 4 \Gamma_R / \Gamma_{R\!\!\!\!/}^c} - 1}{2
\Gamma_R/\Gamma_{R\!\!\!\!/}^c} ,
\ee
where
\be
\Gamma_{R\!\!\!\!/}^c = \frac{1}{16\pi} (.04)^2 M 
\ee
for the favoured scenarios of eqs.(7,8). As we shall see below, 
in these cases one gets a $B \ll$ 1 over a large part of the MSSM
parameter space. The signal for charm and top squark production 
at Tevatron will depend sensitively on this branching fraction. Note 
that for the scenario of eq.(9) one has to replace the factor 
.04 by 0.3 in eq.(13). Consequently the branching fraction B
remains of the order 1 throughout the allowed MSSM parameter space 
in this scenario. 

Let us conclude this section with a brief discussion of the MSSM
parameters, relevant for our analysis [23]. We shall assume a common
gaugino and a common sfermion mass at the unification scale. 
Consequently the masses of the SU(3), SU(2) and U(1) gauginos
($\tilde g$, $\tilde W$ and $\tilde B$) at the electroweak
scale are related via their gauge couplings, i.e.
\beq
M_3 &=& (g^2_s/g^2) \ M_2 \simeq 3.3 M_2 , \nonumber \\ [2mm]
M_1 &=& (5g^{{\prime}{2}}/3g^2) \ M_2 \simeq 0.5 M_2 .
\eeq
Thus there is only one independent gaugino mass $M_2$. Of course 
the electroweak gauginos, $\tilde W$ and $\tilde B$, will mix
with the higgsinos, resulting in the physical chargino ($\tilde 
W_{1,2}$) and neutralino ($\tilde Z_{1-4}$) states. Their masses 
and compositions will depend on two more parameters - the higgsino
mass parameter ($\mu$) and the ratio of the two higgs vacuum expectation
values ($\tan\beta$). Finally, the right and left handed squarks 
of the first two generations are expected to have roughly 
degenerate masses and so also the sleptons. These two masses are 
related via the renormalisation group equations which imply [12]
\be
M^2_{\tilde q} \simeq M^2_{\tilde l} + 0.85 M^2_3 .
\ee
After QCD correction the physical (pole) mass of the gluino is [24]
\be
M_{\tilde g} = (1 + 4.2 \alpha_s/\pi) \ M_3 = 1.15 M_3  \lsim 
M_{\tilde q} .
\ee

Because of the large top quark mass, the top squarks are expected
to have lower masses, with the hierarchy [25]
\be
M_{\tilde t_R} < M_{\tilde t_L} < M_{\tilde q} .
\ee   
Moreover the large top quark mass also implies large mixing between
$\tilde t_{L}$ and $\tilde t_{R}$. This can further reduce the mass
of the lighter physical state
\be
\tilde t_1 = \tilde t_L \cos \theta + \tilde t_R \sin \theta .
\ee
It also implies that $\tilde t_{1}$ has significant left and right
handed components. Thus it is a natural candidate for the anomalous 
HERA events. Indeed the possibility of this top squark production
has been suggested for several years as a promising R-parity
violating SUSY signal at HERA [25]. Therefore, we shall first 
investigate the implications of this scenario for the Tevatron
collider. It should be noted that in this case the $\lambda'$ should
be replaced by $\lambda' cos\theta$ in the second column of Table - I
as well as in eq.(10). However the eqs.(12,13) remain unchanged.

\vskip 2em   

\noindent{\bf 3.~~ \underbar{The Top Squark (Stop) Scenario at
Tevatron}} :\\

The dominant mechanism for stop production at Tevatron are the 
leading order QCD processes of quark-antiquark
and gluon-gluon fusion [26]
\be
%\bar q q \rightarrow \simmee_1 \tilde t_1 \ \ , \ \ gg
%\rightarrow \simmee_1 \tilde t_1 .
\bar q q \rightarrow \tilde{\bar t}_1 \tilde t_1 \ \ , \ \ gg
\rightarrow \tilde{\bar t}_1 \tilde t_1 .
\ee
The R violating Yukawa coupling has negligible contribution to the 
production cross section and hence it is not considered here. Indeed
the above production processes hold equally well for leptoquark
production. It was recently shown in [27] that these LO processes,
combined with the LO structure functions, reproduce the NLO 
cross-section to within 15\%. But combining them with the NLO 
structure functions underestimates the cross-section by a factor
of 1.5 - 2. Therefore we shall use the LO structure functions CTEQ 3L 
[28] in our analysis with the standard choice of the QCD scale Q equal
to the squark mass. We have checked that one gets essentially 
identical results with the GRV 94 LO [29] structure functions.
The structure functions are used via the PDFLIB version 6.06 [30].

Because of the large top mass, $m_t \simeq$ 175 GeV, the stop decay
into the neutralino states can only proceed through higher order
processes,
\be
\tilde t_1 \rightarrow c \tilde Z_1 \ \ {\rm and} \ \ \tilde t_1
\rightarrow b \bar q q' \tilde Z_1 ,
\ee  
as long as 
\be
M_{\tilde t_1} \leq M_{\tilde W_1} .
\ee
The corresponding decay widths are negligibly small [31] compared
to the R violating decay width (13). Therefore, the dominant decay
mode in this case is the R violating decay
\be
\tilde t_1 \rightarrow e^+ d .
\ee
On the other hand the R conserving decay 
\be
\tilde t_1 \rightarrow b \tilde W_i
\ee
will dominate, when kinematically allowed. The corresponding decay
width is given by [25]
\beq
\Gamma (\tilde t_1 \rightarrow b \tilde W_i) &=& {g^2 \over 16\pi}
M_{\tilde t_1} (1 - M^2_{\tilde W_i}/M^2_{\tilde t_1})^2 (A^2_L +
A^2_R) \nonumber \\ [3mm]
A_R &=& - {m_b U_{i2} \cos \theta \over \sqrt 2 M_W \cos \beta} , \
A_L = V^\ast_{i1} \cos \theta + {m_t V^\ast_{i2} \sin \theta \over
\sqrt 2 M_W \sin \beta} ,
\eeq
where $U, V$ are the chargino mixing matrices and we have neglected 
the b mass in the phase space factor.

Fig.1 shows the resulting branching fraction for the R violating 
decay (22) over the relevant MSSM parameter space for $M_{\tilde t_1}$
= 200 GeV. The contours for the lighter chargino mass $M_{\tilde W_1}$
= 85 GeV and 180 GeV are also shown in this figure. The former 
represents the limit of chargino mass which can be probed at LEP-II,
although it has not been done so far for the R violating SUSY model.
We see that the branching fraction for the direct leptonic decay (22)
is $B \leq$ 0.2 for a large part of the parameter space corresponding
to $M_{\tilde W_1} < $ 180 GeV (going up to $M_{\tilde W_1} <
$ 200 GeV for a 220 GeV stop). It should be noted that it corresponds 
to the most favoured region of the MSSM parameter space in terms
of the naturalness criterion [32]. Thus the type of the stop signal  
at Tevatron will be sensitive to the choice of MSSM parameters. Let
us analyse them case by case.

\noindent (A)~~{\bf Direct Leptonic Decays} : This corresponds to the
direct leptonic decays (22) of the stop pair, resulting in a pair of
hard and isolated $e^+ e^-$ along with a pair of jets. Both the
production and decay processes are identical to the leptoquark  case
recently investigated by the CDF and DO$\!\!\!\!/$ collaborations
[33,34], except that in this case there can be no decay into the
neutrino channel. Using a parton level Monte Carlo simulation we have
estimated the signal cross-section following the CDF cuts [33]:
\begin{enumerate}
\item[{(i)}] $E_{Te1}$ and $E_{Te2} > 25$ GeV, $|\eta_{e1}|$ or
$|\eta_{e2}| < 1$, isolation $(E_T^{Ac} < 0.1 E_{Te})$ ;
\item[{(ii)}] $E_{Tj1} > 30$ GeV, $E_{Tj2} > 15$ GeV, $|\eta_{j1, j2}|
< 2.5$, cone size $ R_j = 0.7$;
\item[{(iii)}] $M_{ee} \neq 76 - 106$ GeV;
\item[{(iv)}] $E_{Te1} + E_{Te2} > 70$ GeV, $E_{Tj1} + E_{Tj2} > 70$ GeV.
\end{enumerate}
This is supplemented by the identification efficiency of 0.77 for the
electron pair along with an azimuthal efficiency factor of 0.75
(corresponding to 85\% azimuthal coverage for each electron). We have
checked that the acceptance factor after each set of cuts agrees quite
well with the CDF result. 

Fig.2 shows the signal cross-sections before and after the above
mentioned cuts against the stop mass. The right hand scale shows the
number of events for the integrated luminosity of 110 $pb^{-1}$
corresponding to the CDF data [33]. After these cuts there are 3
remaining events in this data against a SM background of similar size.
The corresponding 95\% CL limit of $\sim$ 8 events implies a stop mass
limit of $\sim$ 200 GeV. One can eliminate the background by suitable
cuts on the $ej$ invariant masses, which can not be implemented
however in our parton level MC simulations. The CDF group has achieved
this while retaining 2/3rd of the signal by requiring the two $ej$
invariant masses to match within 20\% and their average value to agree
with the leptoquark (stop) mass within a $3\sigma$ resolution error.
This is illustrated by the dotted line in Fig.2, which is obtained by
multiplying the long dashed line by 2/3 over the appropriate mass range.
The corresponding 95\% CL limits of 3 events gives a better mass bound
of $M > 210$ GeV. It may be added here that the DO$\!\!\!\!/$ group has
been able to eliminate the background while retaining 80\% of the
signal via a hardness cut
\be
H_T = E_{Te1} + E_{Te2} + E_{Tj1} + E_{Tj2} > 350 GeV.
\ee
Combining this with an optimized set of kinematic cuts and a higher
integrated luminosity of 123 $pb^{-1}$ they have reported a higher
mass limit of $M > 225$ GeV [34]. 

Thus the CDF and DO$\!\!\!\!/$ leptoquark limits would rule out the stop
scenario if the direct leptonic decays are dominant, i.e. $M_{\tilde
W_1} \geq M_{\tilde t_1}$. It may be noted here that the alternative
scenario of eq. (9) corresponds to $B > 0.5$, which may be already
incompatible with the combined CDF and DO$\!\!\!\!/$ data. However the
favoured scenario (8) implies $B \leq 0.2$ over a large part of the
MSSM parameter space corresponding to $M_{\tilde W_1} < 180$ GeV
(Fig.1). This means a reduction of the dilepton signal cross-section
by a factor of at least 25. Evidently the present data sample of CDF
and DO$\!\!\!/$ is in no position to probe a signal of this size. One
expects a 20 fold increase in the integrated luminosity to $2 fb^{-1}$
during the next (Main Injector(MI)) run of Tevatron. Moreover the increase
of the CM energy from 1.8 to 2 TeV corresponds to a 50\% increase in
the cross-section. Thus one expects a 30 fold increase in the signal
size during the MI run, which can probe the stop signal down to $B =
0.2$. Nonetheless there is a significant range of MSSM parameters,
corresponding to $B < 0.2$, which will not be accessible to the
dilepton signal during the MI run. With a further increase of
luminosity to 20 $fb^{-1}$ at TeV33 it may be possible to probe the entire
MSSM parameter space for the stop scenario via the dilepton channel.
Nonetheless it is important to see if one can do better via the other
decay modes of stop.

\noindent (B)~~{\bf Mixed Mode} : A more favourable signature for stop
pair production in the low $B (\lsim 0.2)$ region is provided by the
mixed mode, corresponding to direct leptonic decay (22) of one stop,
while the other undergoes cascade decay
\be
\tilde t_1 \rightarrow b \tilde W_1, \tilde W_1 \rightarrow q' \bar q
\tilde Z_1 , \tilde Z_1 \buildrel{\lambda'_{131}}\over{\rightarrow}
\bar b \bar\nu d + h.c. 
\ee
Note that the R violating decay of the lightest superparticle (LSP) in
this case is restricted to the neutrino channel due to the large top
mass. Thus the final state will consist of a very hard $e^{\pm}$, a
large number of jets including a pair of $b$, and a modest amount of
missing $p_T$ carried by the neutrino. We have estimated the signal
cross-section for the following set of cuts:
\begin{enumerate}
\item[{(i)}] $p^e_T > $ 40 GeV, $|\eta_e| < 1$, isolation $(E^{AC}_T <
0.1 p^e_T)$;
\item[{(ii)}] $n_j \geq 3$ with $E^j_T >$ 15 GeV, $|\eta_j| < 2.5$,
cone size $R_j$ = 0.7;
\item[{(iii)}] $M_T (e,E_T) \neq 50 - 110$ GeV
\item[{(iv)}] $\geq 1 b$-jet with $E^b_T > 20$ GeV and $|\eta_b| < 2$.
\end{enumerate}
The 3rd cut is designed to reduce $W$ decay background. We assume
electron identification and $b$-tagging efficiencies of
\be
\varepsilon_{e-id} = 0.8 \ \, \ \ \  \varepsilon_b = 0.3.
\ee

With the above acceptance cuts and $b$-tagging, the dominant SM
background is expected to come from $\bar t t$ production. Fig.3 shows
the signal cross-section for a 200 GeV stop along with the $\bar t t$
background at $\sqrt s$ = 2 TeV. The MSSM parameters used are
\be
M_2 = 150 GeV, \ \mu = -400 GeV, \ \tan\beta = 2 \Rightarrow M_{\tilde
W_1} = 158 GeV, \ M_{\tilde Z_1} = 77 GeV;
\ee
but the signal cross-section is insensitive to these parameters. The
signal shows a much harder electron $p_T$ distribution than the $\bar
t t$ background. Besides one can exploit the clustering of invariant
mass of the electron with the hardest jet at $M \simeq 200 GeV$, to
distinguish the signal from the background. Thus it seems feasible to
separate the signal from the background while retaining the bulk of
the signal size.

The size of the signal cross-section in Fig.3 is about 50 $fb$,
corresponding to $\sim$ 100 events at the MI run. But it is yet to be
multiplied by the branching factor $\simeq$ 2B. The smallest branching
fraction over the allowed MSSM parameter space (Fig.1) is $B \simeq
7$\%. Hence this signature should be able to probe the stop signal at
the MI run over the full parameter space of MSSM. An interesting
feature of this signal is that the $b$-jet pair will contain unlike as
well as like sign $b$'s with equal probability, though it may be hard
to test it experimentally. 

\noindent (C)~~{\bf Cascade Decays} : The largest event rate for $B
\leq 0.2$ corresponds to the cascade decay (26) of each of the stop
pair. The resulting final state consists of 4 $b$-quarks and a missing
$p_T$ carried by the neutrinos; but the latter is severely degraded
compared to the R conserving case. Consequently the process suffers
from a large background from $\bar t t$ as well as $\bar b b$
production. The leptonic decay of one of the charginos $(\tilde W_1
\rightarrow l \nu \tilde Z_1)$ will give a lepton $(e, \mu)$ 40\% of the
time; but its detectibility will depend on the chargino-neutralino
mass difference. In any case the $\bar t t$ background remains and is
30--40 times larger at the level of the raw cross-section. Therefore
one would need tripple $b$ tag to separate this signal from the
background. We have estimated the signal cross-section for the two
channels with the following cuts:
\begin{enumerate}
\item[{(I)}] ${p\!\!\!/}_T >$ 40 GeV, $n_b \geq 3$ with $E_T^b 
>$ 20 GeV,
$|\eta_b| < 2$;
\item[{(II)}] $p_T^l >$ 15 GeV, $|\eta_l| < 1$, $E_T^{AC} < 0.1 p^l_T$,
 ${p\!\!\!/}_T > 20$ GeV, $n_b \geq 3$ with $E^b_T > 20$ GeV, $|\eta_b| 
< 2$;
\end{enumerate}

assuming a $b$-tagging efficiency $\epsilon_b = 0.3$. Table-II shows the cross-sections for these two channels for a stop
mass of 200 GeV and a CM energy of 2 TeV. The cross-sections are shown
for several values of the $M_2$ and $\mu$ parameters with $\tan\beta =
2$. The cross-sections are similar in size at $\tan\beta = 10$ as
well. The first two rows correspond to the gaugino dominated region
$(M_2 \ll |\mu|)$, characterised by $M_{\tilde Z_1} \simeq M_{\tilde
W_1}/2$. The last two rows correspond to the higgsino dominated region
$(|\mu| \ll M_2)$, characterised by $M_{\tilde Z_1} \simeq M_{\tilde
W_1}$. The first case implies harder lepton $p_T$ but relatively soft
missing-$p_T$, compared to the second. Consequently the missing-$p_T$
signal goes up while the leptonic signal goes down as we go from the
gaugino region to the higgsino one. Notwithstanding this
complementarity, however, the absolute size of the signal is too small
to provide a viable signature for an integrated luminosity of $2
fb^{-1}$. In short, we found no viable signature for detecting stop
pair production, when both of them undergo cascade decay via (26). 

Note that the above conclusion is based on the current b tagging 
efficiency of 30\%. An increase of this efficiency to 50\%, as 
envisaged for the Main Injector run, will result in a 4-5 times 
increase in the signal cross-section. This 
could make it viable over a large range of MSSM parameters.

\vskip 2em

\noindent{\bf 4.~~\underbar{The Charm Squark (Scharm!) Scenario at
Tevatron}} :\\

In this case one expects 8 species of roughly degenerate squarks along
with a gluino of comparable or smaller mass. Only one of them, the
left handed charm squark $\tilde c_L$, has the R violating decay mode
as required to explain the HERA anomaly
\be
\tilde c_L \rightarrow e^+ d .
\ee
In order to estimate its branching fraction $B$, let us note that the
width for the largest R conserving decays into the $\tilde W$
dominated chargino and neutralino states are
\beq
\Gamma (\tilde c_L \rightarrow s \tilde W_i) &=& {g^2 \over 16\pi}
M_{\tilde c}(1-M^2_{\tilde W_i}/M^2_{\tilde c})^2 V^2_{i1},\nonumber\\[2mm]   
\Gamma (\tilde c_L \rightarrow c \tilde Z_i) &=& {g^2 \over 32\pi}
M_{\tilde c} (1 - M^2_{\tilde Z_i}/M^2_{\tilde c})^2 N^2_{i2} .
\eeq
The masses and compositions of these states correspond to 
\be
M_{\tilde W_i, \tilde Z_i} \simeq M_2 \leq {1 \over 3} M_{\tilde c} \
\ {\rm and} \ \ V_{i1}, N_{i2} \simeq 1 .
\ee
Combining these with eqs.(12,13) one sees that the branching fraction
for the R violating decay (29) is 
\be
B \lsim 1/20 .
\ee
Therefore the direct leptonic decay channel (29) will not give a
viable SUSY signature in this case. Instead one has to consider the
cascade decay of the squarks and gluinos into the LSP. Fortunately the
R violating decay of the LSP into $e^\pm$,
\be
\tilde Z_1 \buildrel{\lambda'_{121}}\over{\rightarrow} \bar c e^+ d
(\bar s \bar \nu d) + h.c.
\ee
provides a distinctive like sign dilepton (LSD) signature in this
case. Using this signature one can test this R violating SUSY scenario
over most of the MSSM parameter space even with the present Tevatron
data.

In this case one has to consider a host of production processes [26]
\be
q\bar q (gg) \rightarrow \tilde q \tilde{\bar q}, q \bar q (gg)
\rightarrow \tilde g \tilde g, \ qg (\bar q g) \rightarrow \tilde q
\tilde g ({\tilde{\bar q}} \tilde g) .
\ee
For $M_{\tilde q} < M_{\tilde g}$, the cascade decay proceeds via the
electroweak decays of squarks. Over most of the parameter space of
interest the $\tilde W_1$ and $\tilde Z_2$ states are dominated by the
$\tilde W$ component while the $\tilde Z_1$ is dominated by $\tilde
B$. Thus the dominant decay modes of the left-handed squark are
\be
\tilde q_L \rightarrow q' \tilde W_1 , \ q \tilde Z_2 (\tilde W_1
\rightarrow q \bar q' \tilde Z_1 , \ \tilde Z_2 \rightarrow q \bar q
\tilde Z_1) ,
\ee
while the right-handed squark decays directly into the LSP,
\be
\tilde q_R \rightarrow q \tilde Z_1 .
\ee
For $M_{\tilde g} < M_{\tilde q}$, the cascade decay proceeds via the
3-body decays of gluino, i.e. 
\be
\tilde g \rightarrow q \bar q' \tilde W_1 (50\%) , \ q \bar q \tilde
Z_2 (30\%) , \ q \bar q \tilde Z_1 (20\%) ,
\ee
where the approximate branching fractions are indicated in the bracket
[35]. Finally the produced pair of LSP undergo the R violating decay
(33).

Thanks to the Majorana nature of the LSP $(\tilde Z_1)$, the final
state di-electron have equal probability of having unlike and like
signs. The latter constitutes a viable signature due to the low SM
background in this channel. Indeed the LSD signature for R violating
SUSY model has been considered in [35,36] for a variety of the
$\lambda$ and $\lambda'$ couplings. We shall concentrate here on
$\lambda'_{121}$ as the leading R violating Yukawa coupling, as
suggested by the charm squark scenario. We have estimated the LSD
cross-section for the following kinematic cuts:
\be
p^l_T > 15 GeV , \ |\eta_l| < 1 , \ \ {\rm Isolation} \ \ (E^{AC}_T <
5 GeV)\ \ .
\ee
The SM background for these cuts has been estimated to be only 2.4
$fb$ at the CM energy of 1.8 TeV, coming mainly from WZ and $t \bar t$
channels[36].

Fig.4 shows the LSD signal cross-section at a CM energy of 1.8 TeV for
different choices of the MSSM parameters. In each case the signal is
shown for a common squark mass $M_{\tilde q} = $ 210 GeV, with
$M_{\tilde g} =$ 150--240 GeV. The upper limit of $M_{\tilde g}$ is
suggested by eqs.(15,16)[37]. The contributions from the three finals
states of eq.(34) to the signal are shown separately. They include the
contributions from the small leptonic components in $\tilde W_1, \tilde
Z_2 \rightarrow \tilde Z_1$ decays. But the bulk of the contribution 
corresponds
to the dielectron coming from the LSP decays. We have not included any
efficiency factor for electron identification. But it is clear from
this figure that, even after making allowance for the identification
efficiency, one expects to see at least half a dozen like sign
dielectron events with the present Tevatron luminosity of $110 pb^{-1}$. We
conclude this section with the hope that this data will be analysed
soon to probe the R violating SUSY model, and in particular to test
the charm squark scenario for the anomalous HERA events. Indeed it should be
possible to do this with the dilepton data even without charge identification,
where one can controll the SM background by choosing suitable kinematic cuts.

\vskip 2em

\noindent{\bf 5.~~\underbar{Summary}} :\\

The two favoured scenarios for the anomalous HERA events in the
R-parity violating SUSY model are the production of top and charm
squarks off the valence $d$ quark via the Yukawa couplings
$\lambda'_{131}$ and $\lambda'_{121}$ respectively. We have studied
the prospects of testing these scenarios at the Tevatron collider
assuming MSSM with common superparticle masses at the unification
scale. In this case the size of the required $\lambda'$ coupling and
the corresponding decay branching fraction $B$ can be independently
estimated as functions of the MSSM parameters. We find $B \ll 1$ for
the charm squark, while the same is also true for the stop over a
large range of the MSSM parameters. Consequently the canonical
leptoquark signature of dilepton plus dijets is of limited value in
probing these scenarios at Tevatron. We suggest alternative
signatures, which seem to be viable over the entire parameter space of
interest. We have also considered the alternative scenario for stop
production from a strange quark via the $\lambda'_{132}$ coupling. In
this case $\lambda' > 1/2$; and this scenario may be already in
conflict with the combined Tevatron data via the leptoquark signature.

We gratefully acknowledge discussions with Drs. V. Barger, A.S. Belayev, D. Choudhury, S. Eno,
S. Majumdar, N.K. Mondal, N. Parua, K. Sridhar and G. Wang. 

\newpage

\noindent{\bf REFERENCES} :

\begin{enumerate}
\item[{1.}] H1 Collaboration: C. Adloff et.al, {\it Z. Phys.} {\bf C74}, 191
(1997).
\item[{2.}] ZEUS Collaboration: J. Breitweg et.al, {\it Z. Phys.} {\bf C74},
207 (1997).
\item[{3.}] G. Wolf, hep-ex/9704006; see however M. Drees, hep-ph/9703332.
\item[{4.}] D. Choudhury and S. Raychaudhuri, hep-ph/9702392 and
hep-ph/9703369.
\item[{5.}] H. Dreiner and P. Morawitz, hep-ph/9703279 (version 2).
\item[{6.}] G. Altarelli, J. Ellis, G.F. Giudice, S. Lola and M.L. Mangano,
hep-ph/9703276. 
\item[{7.}] A.S. Belayev and A.V. Gladyshev, hep-ph/9703251 and
hep-ph/9704343; J. Kalinowski, R. Ruckl, H. Spiesberger and P.M.
Zerwas, hep-ph/9703288.
\item[{8.}] K.S. Babu, C. Kolda, J. March-Russell and F. Wilczek,
hep-ph/9703299. 
\item[{9.}] V. Barger, K. Cheung, K. Hagiwara and D. Zeppenfeld,
hep-ph/9703311. 
\item[{10.}] J.L. Hewett and T. Rizzo, hep-ph/9703337. 
\item[{11.}] G.G. Ross, Grand Unified Theories, Benjamin Cummings
(1985); J.L.Hewett and T. Rizzo, {\it Phys. Rep.} {\bf 183}, 193(1989).
\item[{12.}] H.P. Nilles, {\it Phys. Rep.} {\bf 110}, 1 (1984); R.
Arnowitt, A. Chamseddine and P. Nath, Applied $N=1$ Supergravity,
World Scientific (1984); H. Haber and G. Kane, {\it Phys. Rep.} {\bf
117}, 75 (1985); A.B. Lahanas and D.V. Nanopoulos, {\it Phys. Rep.}
{\bf 145}, 1 (1987).
\item[{13.}] S. Weinberg, {\it Phys. Rev.} {\bf D26}, 287 (1982); N.
Sakai and T. Yanagida, {\it Nucl. Phys.} {\bf B197}, 83 (1982); S.
Dimopoulos, S. Raby and F. Wilczek, {\it Phys. Lett.} {\bf B212}, 133
(1982). 
\item[{14.}] M. Hirsch, H.V. Klapdor-Kleingrothaus and S.G. Kovalenko,
{\it Phys. Rev. Lett.} {\bf 75}, 17 (1995); {\it Phys. Rev.} {\bf
D53}, 1329 (1996).
\item[{15.}] V. Barger, G.F. Giudice and T. Han, {\it Phys. Rev.} {\bf
D40}, 2987 (1989).
\item[{16.}] S. Davidson, D. Bailey and B. Campbell, {\it Z. Phys.}
{\bf C61}, 613 (1994); P. Langacker, {\it Phys. Lett.} {\bf B256}, 277
(1991). 
\item[{17.}] R.M. Godbole, P. Roy and X. Tata, {\it Nucl. Phys.} {\bf
B401}, 67 (1993).
\item[{18.}] J. Ellis, S. Lola and K. Sridhar, hep-ph/9705416.
\item[{19.}] C. S. Wood et. al, Science {\bf 275}, 1759(1997).  
\item[{20.}] P. Langacker and J. Erler, Standard Model of Electrowek
Int., in Review of Particle properties, M. Barnett et al, {\it Phys 
Rev} {\bf D54}, 1(1996).
\item[{21}] G. Altarelli, G. F. Giudice and M. L. Mangano, hep-ph/
9705287.
\item[{22.}] Z. Kunszt and W.J.Stirling, hep-ph/9703427; T.Plehn, H.
Spiesberger, M. Spira and P.M. Zerwas, hep-ph/9703433.
\item[{23.}] For a micro-review see H. Haber, Super Symmetry, in Review
of Particle Properties, M. Barnett et.al, {\it Phys. Rev.} {\bf D54}, 1
(1996). 
\item[{24.}] S.P. Martin and M.T. Vanghn, {\it Phys. Lett.} {\bf
B318}, 331 (1993); N.V. Krasnikov, {\it Phys. Lett.} {\bf B345}, 25 (1995).
\item[{25.}] T. Kon and T. Kobayashi, {\it Phys. Lett.} {\bf B270}, 81
(1991); T. Kon, T. Kobayashi and S. Kitamura, {\it Phys. Lett.} {\bf
B333}, 263 (1994); {\it Int. J. Mod. Phys.} {\bf A11}, 1875 (1996).
\item[{26.}] G.L. Kane and J.P. Leveille, {\it Phys. Lett.} {\bf 
B112}, 227 (1982); P.R. Harrison and C.H. Llewellyn-Smith, {\it Nucl.
Phys.} {\bf B213}, 223 (1983) [Err. {\it Nucl. Phys.} {\bf B223}, 542
(1983)]; S. Dawson, E. Eichten and C. Quigg, {\it Phys. Rev.} {\bf
D31}, 1581 (1985); E. Reya and D. P. Roy, {\it Phys. Rev.} {\bf D32}
, 645 (1985).     
\item[{27.}] M. Kramer, T. Plehn, M. Spira and P.M. Zerwas, hep-ph/9704322.
\item[{28.}] H.L. Lai et.al, {\it Phys. Rev.} {\bf D51}, 4763 (1995).
\item[{29.}] M. Gluck, E. Reya and A. Vogt, {\it Z. Phys.} {\bf C67},
433 (1995). 
\item[{30.}] PDFLIB, Version 6.06, H. Plothow-Besch, CERN-PPE (1995-03.15).
\item[{31.}] K. Hikasa and M. Kobayashi, {\it Phys. Rev.} {\bf D36},
724 (1987).
\item[{32.}] J. Ellis, K. Enqvist, D.V. Nanopoulos and F. Zwirner,
{\it Mod. Phys. Lett.} {\bf A1}, 57 (1986); R. Barbieri and G.F.
Giudice, {\it Nucl. Phys.} {\bf B306}, 63 (1988); G.W. Anderson and D.J. Castano, {\it Phys. Rev. } {\bf D52}, 1693 (1995). 
\item[{33.}] CDF Collaboration: C. Grosso-Pilcher, Talk at the
Vanderbilt Conference, May 1997.
\item[{34.}] DO$\!\!\!\!/$ Collaboration: D. Norman, Talk at the Hadron
Collider Physics Conference, Stonybrook, June 1997.
\item[{35.}] M. Guchait and D.P. Roy, {\it Phys. Rev.} {\bf D54}, 3276 (1996).
\item[{36.}] H. Baer, C. Kao and X. Tata, {\it Phys. Rev.} {\bf D51},
2180 (1995).
\item[{37.}] For $M_{\tilde g} <$ 240 GeV the positive values of
$\mu$ are disfavoured by the LEP bounds on $\tilde W_1$ and 
$\tilde Z_1$ masses[35]. The same is true for $\mu =$--100 GeV at
$tan\beta =$ 10. Therefore we do not include these points in fig.4.
\end{enumerate}

\newpage
\begin{center}
{\bf Table--I}: Different squark formation and decay processes are
shown along with the size of the corresponding R violating Yukawa
couplings, required to explain the HERA events. The last column shows
the upper limits on these couplings from other processes.

\begin{tabular}{|lcl|}\hline 
Process & Reqd. $\lambda' \sqrt B$ & $\lambda'$ limit $(M_{\tilde q}
\simeq$ 200 GeV) \\ \hline
$e^+ d \rightarrow \tilde u_j \rightarrow e^+ d$ & 0.04 &
$\lambda'_{111} < .004 , \ \lambda'_{121,131} < .13(.10)$ \\
&& \\
$e^+ s \rightarrow \tilde u_j \rightarrow e^+ s$ & 0.3 &
$\lambda'_{112} < .06 , \ \lambda'_{122} < .09 , \ \lambda'_{132} < 0.6$ \\
&& \\
$e^+ b \rightarrow \tilde u_j \rightarrow e^+ b$ & 0.6 &
$\lambda'_{113} < .06 , \ \lambda'_{123} < .55 , \ \lambda'_{133} < .003$ \\
&& \\
$e^+ \bar u \rightarrow {\tilde {\bar d}_k} \rightarrow e^+ \bar u$ & 0.3 &
$\lambda'_{111} < .004 , \ \lambda'_{112,113} < .06$ \\
&& \\
$e^+ \bar c \rightarrow {\tilde {\bar d}_k} \rightarrow e^+ \bar c$ & 0.4 &
$\lambda'_{121} < .13 , \ \lambda'_{122} < .09 , \ \lambda'_{123} <
.55$ \\ \hline
\end{tabular}

\vskip 3em

{\bf Table--II}: Stop cross-section in the cascade decay channels with
and without a lepton for $M_{\tilde t_1}$ = 200 GeV, $\sqrt s$ = 2 TeV
and the cuts described in the Text. The results are shown for several
gaugino and higgsino mass parameters with $\tan\beta$ = 2.

\begin{tabular}{|cccc|cc|} \hline
$M_2$ & $\mu$ & $M_{\tilde W_1}$ & $M_{\tilde z_1}$ & $\sigma (fb)$ \\
(GeV) & (GeV) & (GeV) & (GeV) &\multicolumn{2}{c} (I) ${p\!\!\!/}_T 
+ \geq 3b \
(II) l + {p\!\!\!/}_T + \geq 3b$ \\ \hline
150 & --400 & 158 & 77 & 3.3 & 2 \\
100 & --800 & 105 & 51 & 2 & 1.4 \\
100 & --100 & 101 & 51 & 3 & 1.5 \\
200 & --100 & 112 & 89 & 6 & 1.2 \\
300 & --100 & 111 & 95 & 8 & 0.8 \\ \hline
\end{tabular}
\end{center}

\vspace{5cm}

\noindent\underbar{\large{\bf Figure Captions }}\\
\smallskip
\begin{enumerate}
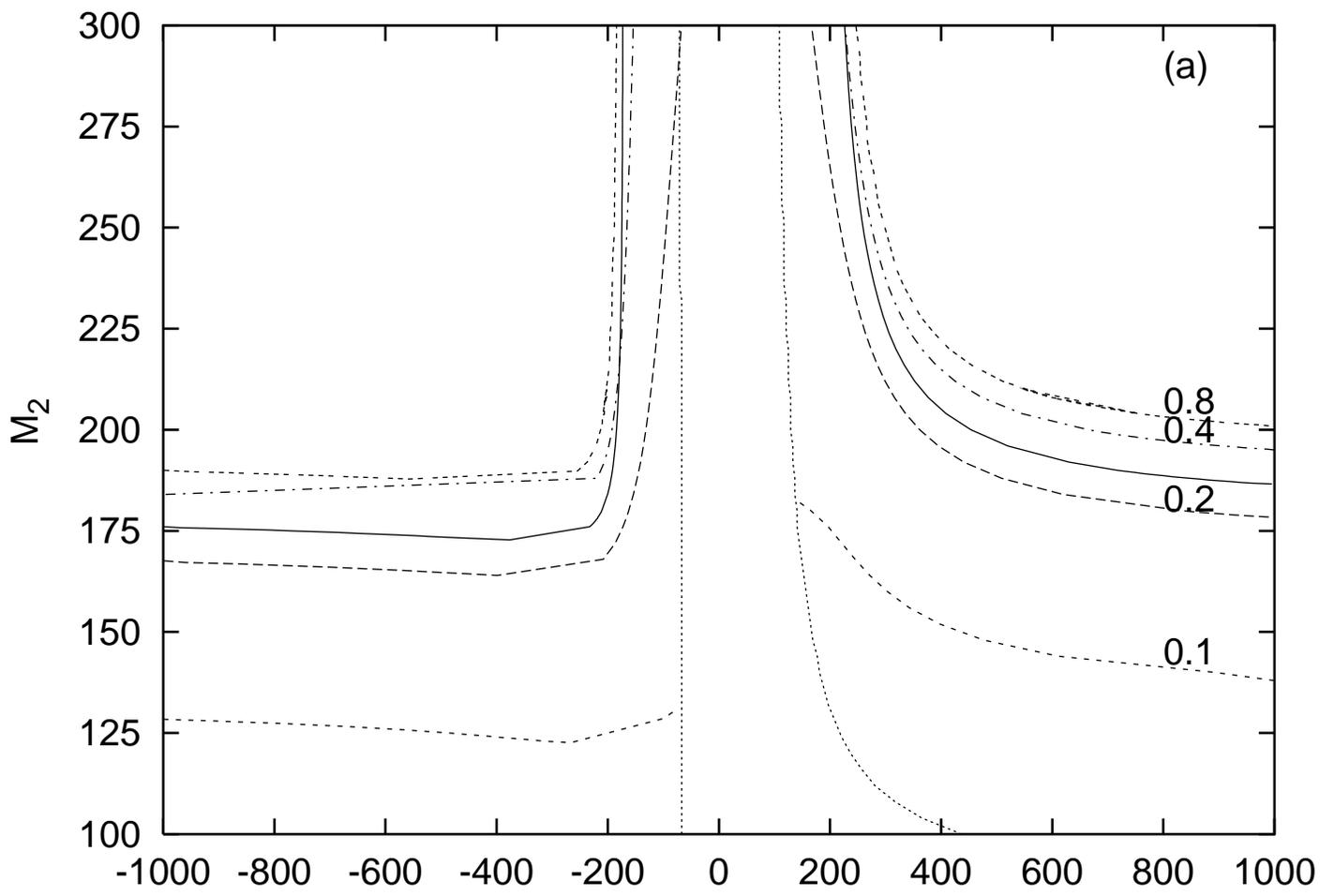
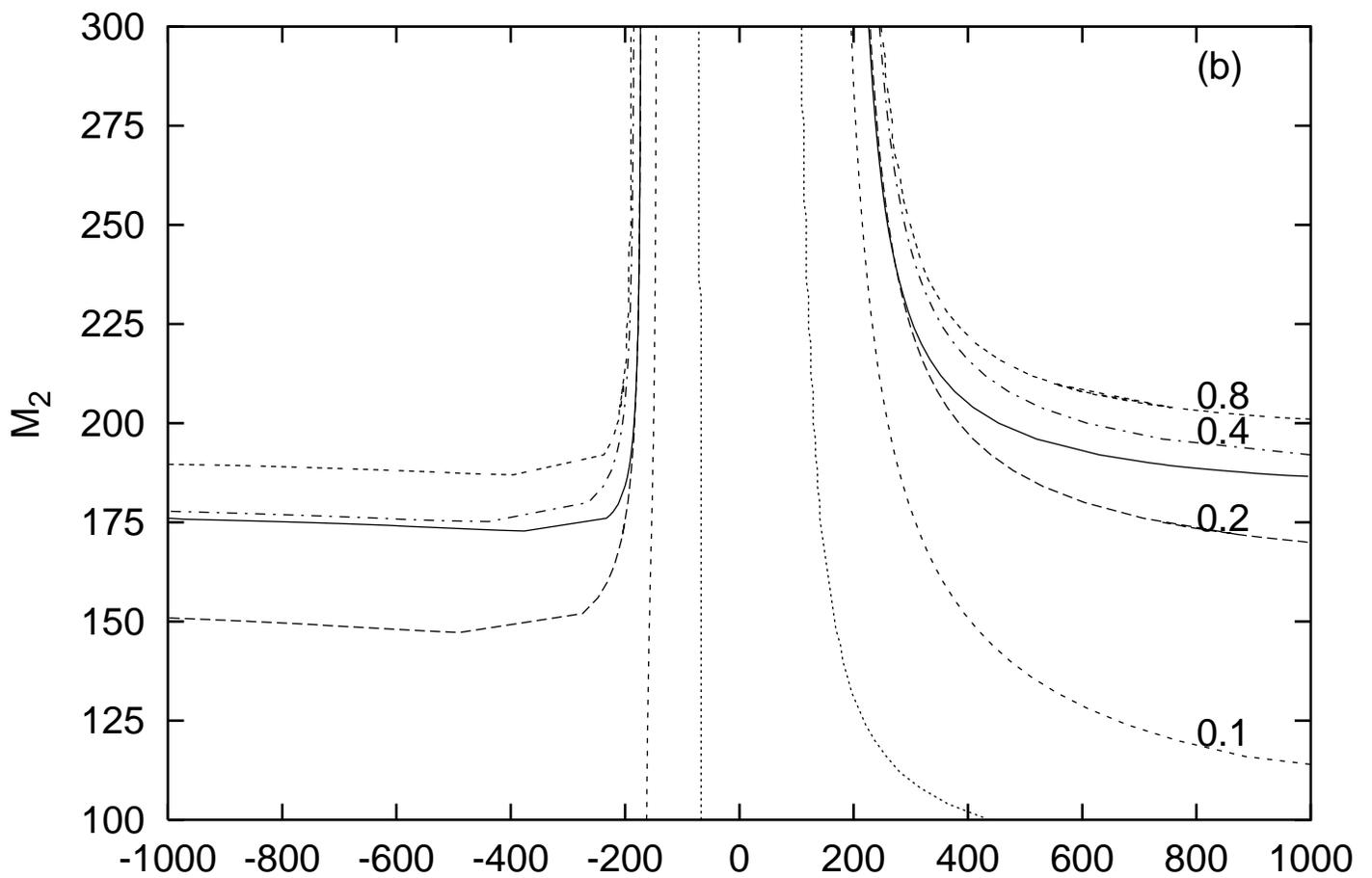
\item[{Fig.1}]: The stop branching fraction(eq.11) is shown as a contour plot in the
$M_2$, $\mu$ plane for the stop mass $M_{\tilde t_1}= 200 GeV$, 
$tan\beta = 2$ and 
(a) $\theta_
{\tilde t} = 0^\circ$ (b) $\theta_{\tilde t} = -45^\circ$.
Contours for the lighter chargino mass of 180GeV(solid line)
and 85GeV(dotted line) are shown.

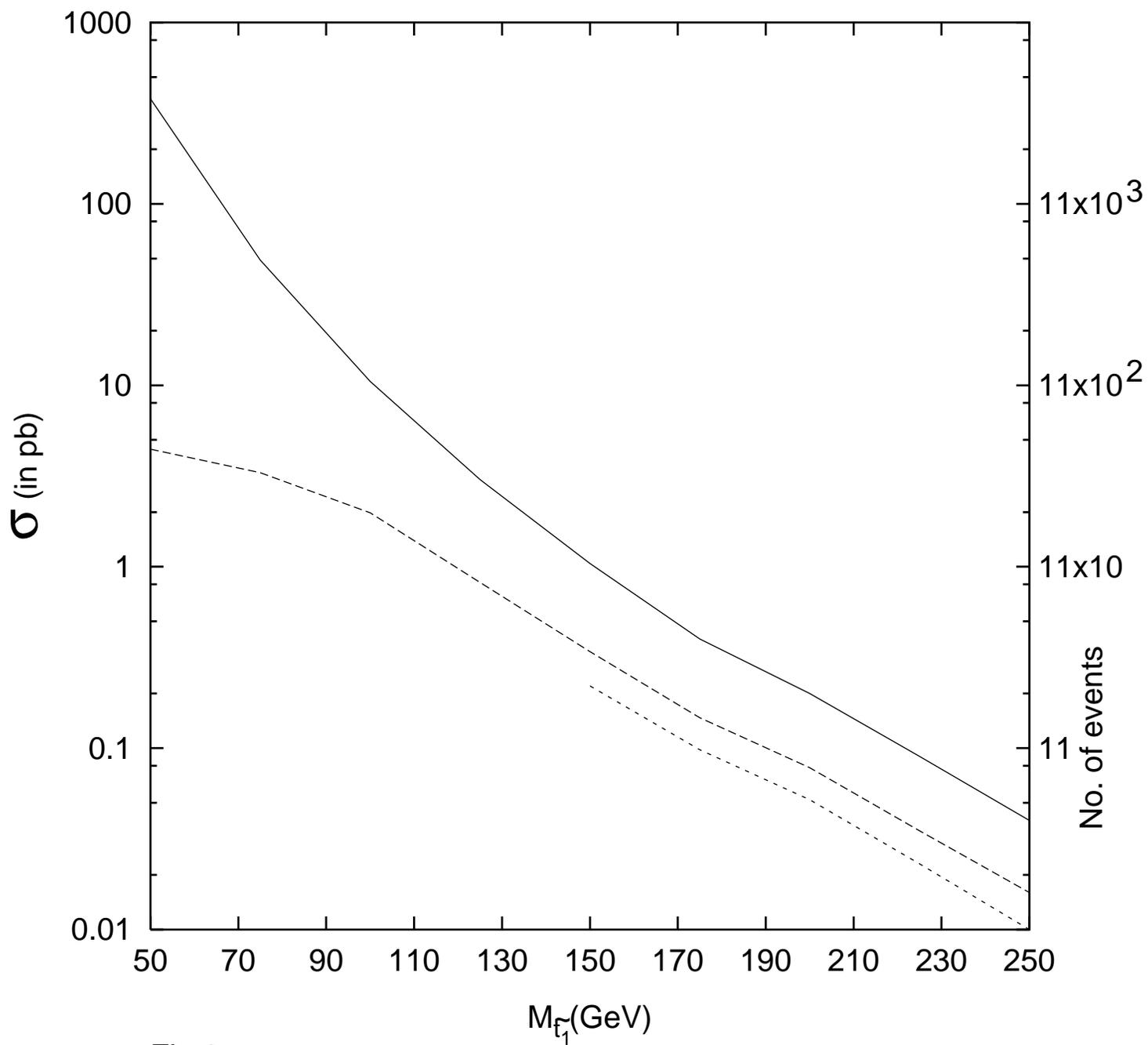
\item[{Fig.2}]:
The stop pair-production cross section(No. of events for 110 $pb^{-1}$ luminosity) shown 
against the stop mass at the Tevatron collider energy of 1.8 TeV. The solid line corresponds to the 
raw cross section, while the long and
short dashed lines correspond
to the dielectron+dijet cross sections following the CDF cuts as described in the text. 

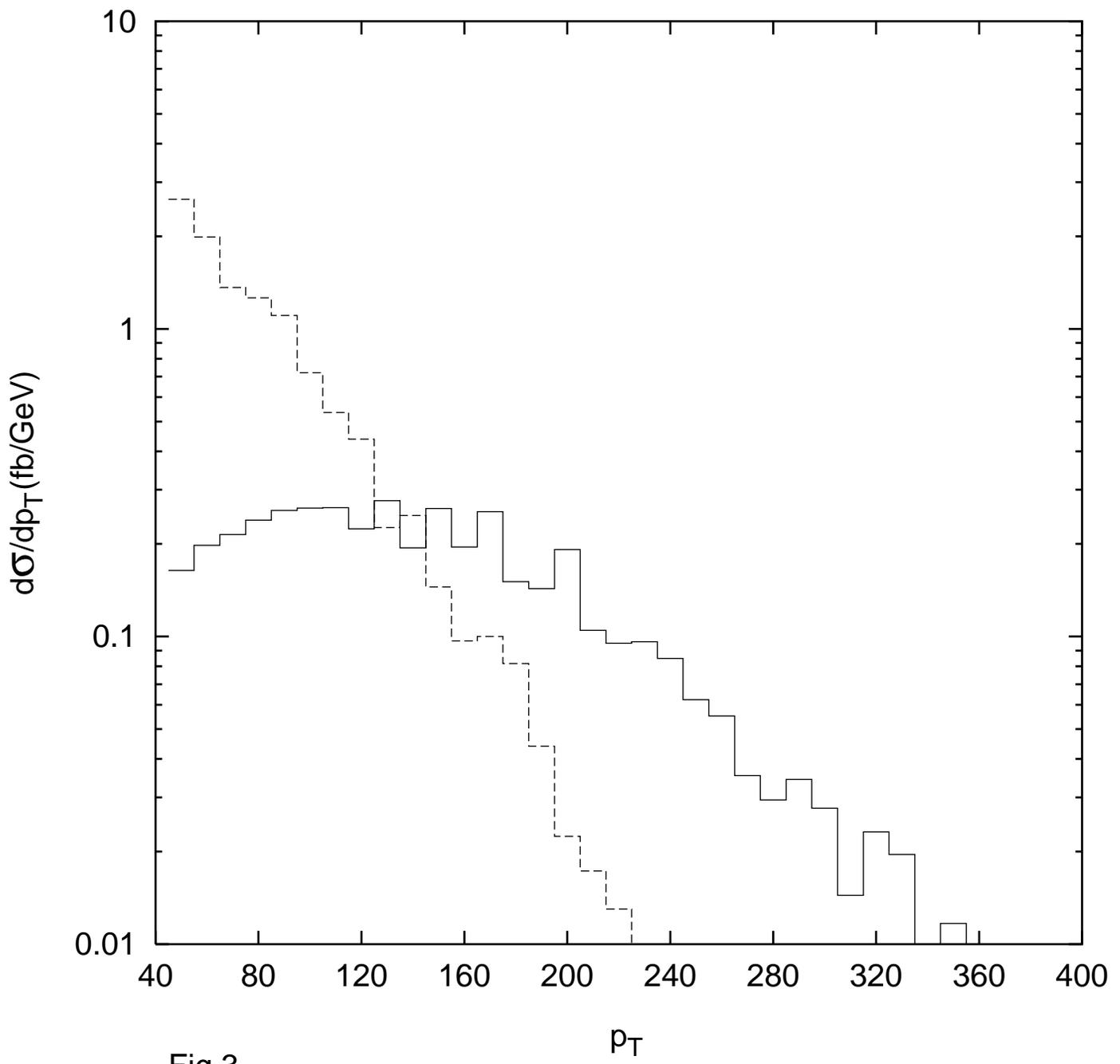
\item[{Fig.3}]:
The signal cross section(solid line) corresponding to the mixed 
mode(B) is shown for stop mass of 200 GeV along with the $t \bar t$
background(dashed line)  
against the $p_T$ of the electron at 2 TeV. The MSSM parameters are $M_2 = 
150 GeV$,
$\mu = -400 GeV$, and $tan\beta = 2$. 
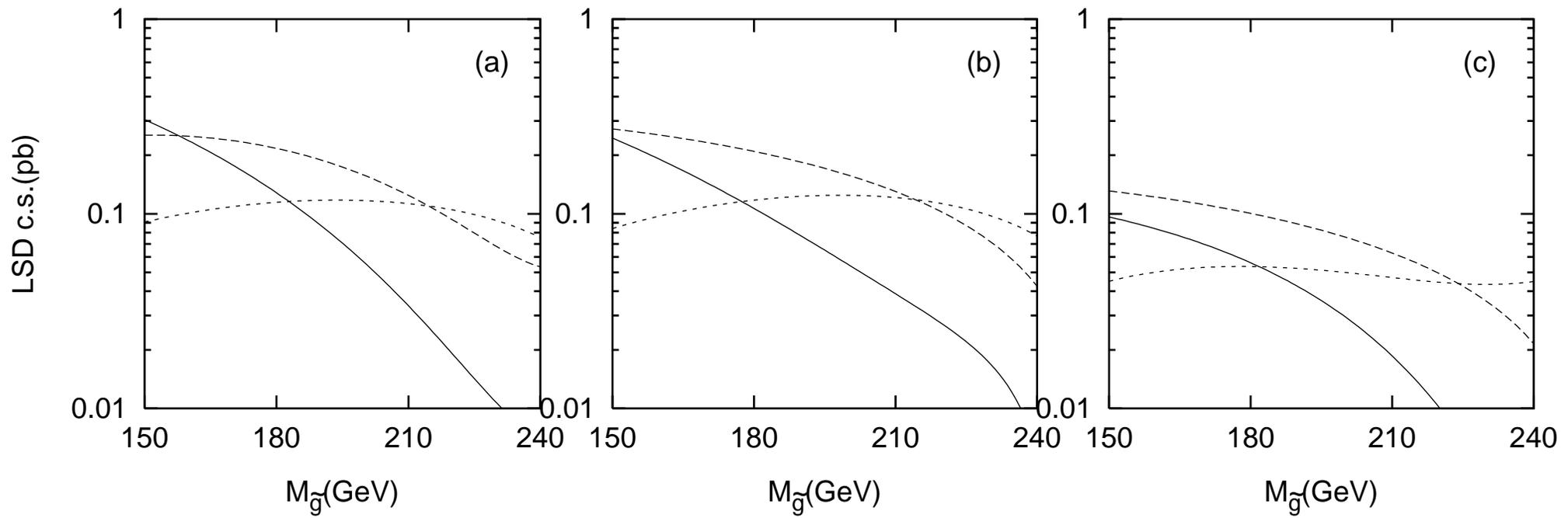
\item[{Fig.4}]:
The gluino-gluino(solid line), squark-antisquark(short dashed) and squark-gluino(long
dashed) contributions to LSD cross section shown as a function
of gluino mass for ($\mu$,$tan\beta$) =  (a) --100 GeV, 2 (b) --300 GeV, 2
(c)--300 GeV, 10.  
\end{enumerate}
\end{document}